# The International Linear Collider

Jim Brau[†], Paul Grannis[‡], Mike Harrison[#], Michael Peskin[*], Marc Ross[*], Harry Weerts[§]
for the ILC Collaboration
April 9, 2013

The motivation for the ILC is driven by important physics goals for the TeV energy scale: the need to measure precisely the properties of the newly discovered Higgs-like boson, including its couplings to fermions and bosons, the need to bring our knowledge of the top quark to a high level of precision, and the need to pursue possible signals of new physics through the electroweak production of new particles and through signals of these interactions in W, Z, and two-fermion processes.  The ILC experiments will be sensitive to phenomena such as supersymmetric partners of known particles, new heavy gauge bosons, extra spatial dimensions, and particles connected with strongly-coupled theories of electroweak symmetry breaking [1].  In all of these sectors, the ILC will yield substantial improvements over LHC measurements.  Knowledge of Higgs boson couplings at the few percent level is needed to determine whether this object is that expected in the Standard Model, if it arises from new physical mechanisms, or if it couples to new particles inaccessible in other ways.  Detailed simulations with realistic detector designs show that the ILC can reach this precision.  While we recognize that the LHC experiments are now making more precise measurements than were originally predicted (as was also the case with the Tevatron, LEP and SLC experiments), we should also expect that ILC experiments will bring qualitatively new capabilities and will similarly exceed the performance levels based on simulations when data are in hand.

The high level parameters of the ILC were established in 2003 [2] for a machine that can be tuned to run between 200 and 500 GeV, and is capable of rapid changes in energy over a limited range for threshold scans.  The luminosity required should exceed $10^{34}$ cm$^{-2}$s$^{-1}$ at 500 GeV, roughly scaling in proportion to the collision energy.  The key characteristics of the ILC accelerator are the relatively long interval between collisions of bunches (allowing localization of signals to a specific bunch crossing), narrow beam energy spread, beam position and energy stability, and the ability to polarize both electrons and positrons.  The TDR design [3] meets these specifications.  In a staged approach starting with 250 GeV $e^+e^-$ operation for the Higgs boson study, it should be possible to reach the physics goals for Higgs branching ratios and properties with about five years of operation, including an initial ramp up to full luminosity.  Raising the energy to 500 GeV will allow precision measurements of the top quark mass and its properties well beyond those possible at the LHC and Tevatron. Measurements of the top coupling to the Higgs and the Higgs self coupling would begin at 500 GeV.  Should there be accessible new particles such as supersymmetric partners of gauge bosons and leptons, the ILC is the only place where they can be studied in full detail. If there are multiple Higgs bosons, the ILC would be needed to measure their branching fractions and the mixing angle tanβ. Further details of the Higgs spectrum, such disentangling the nearly-degenerate heavy CP even and odd Higgs particles expected in supersymmetric models, could be achieved with polarized γγ collisions of high enough energy. The γγ option can be installed in the ILC as designed with the addition of the required high power lasers to induce Compton backscattered photons of about 80% of the incoming electron/positron beam energies. The ILC could be operated as an $e^-e^-$ collider if there is a physics need.

The 500 GeV ILC technical design presented in the ILC Technical Design Report (TDR) [3] has been developed based on an extensive R&D program [4]. The R&D has successfully demonstrated the goal of 35 MV/m accelerating gradients in test stands and 31.5 MV/m in installed cryomodules with beam loading, using niobium cavities with no more than two surface-preparation processing cycles. Cavity fabrication to these specifications has been industrialized, with qualified vendors in Europe, North America, and Asia. With this accelerating gradient, the total length of the 500 GeV ILC is 31 km. The effects of the electron cloud in the positron damping ring have been studied experimentally, leading to the proven techniques included in the TDR design for its mitigation. The fast kickers needed to inject and eject beams from the damping rings have been developed. The ability to achieve the small final focus spot size is being demonstrated in a test facility and gives confidence that the goal of several nm vertical spot sizes will be achieved. The final focus and interaction region, including the detector push-pull system needed to allow two detectors to take data sequentially, has been designed. The TDR design and the R&D results have been judged sufficient [5] to begin the detailed, site specific design and construction stage once international negotiations for starting the project have been concluded. Remaining work includes beam tests in multi-cryomodule facilities now under construction to assess such topics as beam stability, low level RF controls and field emission behavior; as well as further industrialization of SCRF cavity and cryomodule components, value engineering, and detailed site-specific engineering design.

In parallel with the accelerator design and technology TDR , two detailed detector designs have been developed [6]. The R&D to support these designs has advanced to a high level of maturity. For example, beam tests with highly granular calorimeters have demonstrated the needed performance of calorimetry based on the particle flow technique. Similarly, tracking R&D has advanced for vertex detection based on thin CMOS monolithic pixel sensors, outer tracking with low-mass supported silicon microstrips, and advanced TPC technologies employing micropattern gas detectors or silicon sensors for readout.

Extension of the ILC to 1 TeV is straightforward. It requires lengthened linac tunnels and additional cryomodules but will use the original ILC sources, damping rings, final focus and interaction regions, and beam dumps. No new technological breakthroughs would be required, although R&D to develop higher gradient cavities would permit shorter tunnel extensions and thus cost savings. The power required for the ILC operation at 500 GeV is 162 MW, including the emergency power systems required in case of main power failure. For the 250 GeV Higgs factory option, a high luminosity 500 GeV option, and the 1 TeV upgrade the power required is 128 MW, 205 MW, and 300 MW, respectively.

Superconducting RF techniques have wide applicability in other science facilities. The technical development for the ILC is already applied in X-ray light sources and spallation neutron sources, and it has enabled the design of energy recovery linacs. The SCRF technology has attracted considerable interest for military and homeland security systems [7]. The high-precision goals of the ILC have driven the development of new detector technologies such as high-density low-power silicon pixels, compact calorimeters with data readout and sparsification using on-detector ASICs, and new algorithms using particle flow techniques and digital sampling. These advances will influence a broad range of future experiments [8].

The cost estimate for the 500 GeV ILC has been produced using a 'value estimate' methodology[9]. This is the norm for large-scale internationally funded projects that are constructed using in-kind contributions. The value estimate for the construction of the ILC is 7.8 billion ILC Units together with 23 million person hours (approximately 13,000 person years) of additional labor, (one ILC Unit is equivalent to one 2012 USD) [10]. The estimate covers all construction costs for the accelerator complex. It does not include contingency and escalation, commissioning with beam or operations costs. It also does not include costs for project engineering, site acquisition and preparation costs, costs of R&D prior to construction start, and the cost of the detectors.

The GDE has not as yet produced an optimized design and cost for a reduced scope machine running at 250 GeV. The cost savings based on the current value estimate for the 500 GeV machine would be in the range of 25-30%. Similarly, doubling the center-of-mass collision energy will cost 65 to 75% more than the 500 GeV ILC baseline.

**References**

† University of Oregon, ‡ Stony Brook University, # Brookhaven National Laboratory
* SLAC National Accelerator Laboratory, § Argonne National Laboratory
(The full list of signatories to the 2013 ILC Technical Design Report will be available in June 2013.)

1. *Physics at the International Linear Collider*, http://ific.uv.es/~fuster/DBD-Chapters/ (2013).

2. *Parameters for the Linear Collider*, http://www.fnal.gov/directorate/icfa/recent_lc_activities_files/para-Nov20-final.pdf (2003, updated 2006).

3. Technical Design Report, *The ILC Baseline Design*, https://forge.linearcollider.org/dist/20121210-CA-TDR2.pdf (2013).

4. Technical Design Report, *ILC R&D in the Technical Design Phase*, https://forge.linearcollider.org/dist/20121210-CA-TDR1.pdf (2013).